\documentclass[aps,prl,twocolumn,showpacs,groupedaddress,letterpaper]{revtex4}
\usepackage{graphicx}
\usepackage{amssymb}
\usepackage{amsmath}
\newlength{\figwidth}

\begin{document}
\preprint{\today}
\title{Thermal Enhancement of Interference Effects in Quantum Point Contacts}

\author{Adel Abbout}
\affiliation{Service de Physique de l'\'Etat Condens\'e (CNRS URA 2464), 
IRAMIS/SPEC, CEA Saclay, 91191 Gif-sur-Yvette, France}

\author{Gabriel Lemari\'e}
\affiliation{Service de Physique de l'\'Etat Condens\'e (CNRS URA 2464), 
IRAMIS/SPEC, CEA Saclay, 91191 Gif-sur-Yvette, France}

\author{Jean-Louis Pichard}
\affiliation{Service de Physique de l'\'Etat Condens\'e (CNRS URA 2464), 
IRAMIS/SPEC, CEA Saclay, 91191 Gif-sur-Yvette, France}

\begin{abstract} 
 We study an electron interferometer formed with a quantum point contact 
and a scanning probe tip in a two-dimensional electron gas. The images 
giving the conductance as a function of the tip position exhibit fringes 
spaced by half the Fermi wavelength. For a contact opened at the edges of 
a quantized conductance plateau, the fringes are enhanced as the temperature 
$T$ increases and can persist beyond the thermal length $l_T$. This unusual 
effect is explained assuming a simplified model: The fringes are mainly given 
by a contribution which vanishes when $T \to 0$ and has a decay characterized 
by a $T$-independent scale.
\end{abstract}

\maketitle

\begin{figure}[b]
\begin{center}
\includegraphics[height=4.2cm]{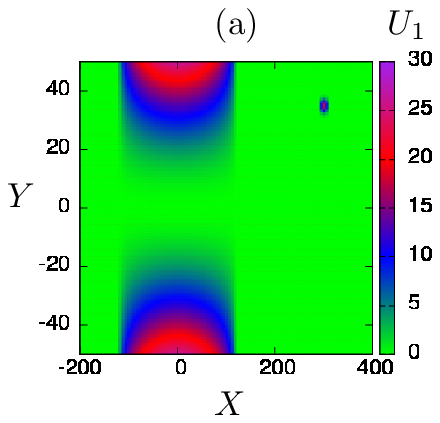}
\includegraphics[height=4.2cm]{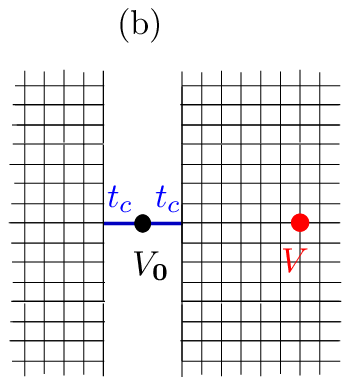}
\end{center}
\caption{(color online) Electron interferometer made of 
a QPC and a scatterer (red point on the right) induced by a charged tip.
Fig.~(a): Saddle-point QPC potential used in Model 1: 
$U_{1}(X,Y)= \left(Y/L_Y\right)^2 \left[1-3(X/L_X)^2+2 |X/L_X|^3 \right]^{1/4}$ for $ \vert X \vert \leq L_X$ 
and $U_1=0$ elsewhere. $L_Y=10$ and $L_X=120$ (i.e.~long QPC). Model 2 (hard wall QPC not shown) consists in 
removing for $ \vert X \vert \leq L_X$ the sites of coordinates $|Y|\geq L_Y + 3\left(X/L_X\right)^2$, 
with $L_Y=6$ and $L_X=20$ but keeping otherwise the site potentials equal to zero. The hopping amplitudes 
are $t_h=-1$. Fig.~(b): Resonant Level Model (RLM): 2 semi-infinite square lattices (with zero on-site potential 
and hopping amplitudes $t_h=-1$) are contacted by hopping terms $t_c$ via a single site $\bf{0}$ of energy 
$V_{\bf{0}}$. At a distance $x$ from the contact, there is a tip potential $V\neq 0$.}
\label{fig1}
\end{figure}
The quantum circuits used in nanoelectronics are very often built in a two-dimensional electron gas (2DEG)
made of a thin sheet of conduction electrons created just beneath the surface of a semiconductor heterostructure. 
Induced by electrostatic gates deposited at the surface of the heterostructure, the quantum point contact (QPC) 
is one of the elementary elements of quantum circuits used in a wide variety of investigations, including transport 
through quantum dots, Mach-Zehnder interferometry and various prototypes of quantum-computing schemes.
The quantization~\cite{PhysRevLett.60.848,Wharam} in units of $2 e^2/h$ of its conductance $g$ can be explained 
using simple non-interacting models~\cite{Glazman:JETP88, Buttiker:PRB90, Beenakker:SSP91}, outside some anomalies, 
as the $0.7 (2e^2/h)$ anomaly~\cite{PhysRevLett.77.135}, which cannot be explained by a non interacting theory.
The recent engineering~\cite{Topinka:Sci00, Topinka:Nat01, Topinka:PT03, LeRoy:PRL05} 
of the scanning gate microscope (SGM) has allowed to ``image'' the electron flow associated with the successive 
conductance plateaus \cite{Topinka:Sci00, Topinka:PT03} of a QPC. The images are obtained with the charged tip of an 
AFM cantilever which can be scanned over the surface of the heterostructure. A negatively charged tip causes a depletion 
region in the 2DEG underneath the tip which scatters the electrons at a distance $r$ from the QPC. The tip and the 
QPC form an electron interferometer, and the SGM images give its conductance $g$ as a function of the tip position. 
Fringes falling off with $r$ and spaced by half the Fermi wavelength $\lambda_F/2$ characterize these images.

 In mesoscopic physics, the interference effects are usually important when the temperature $T=0$, and 
disappear as $T$ increases at scales larger than the thermal length $l_T \propto 1/T$. We discuss here the 
possibility to observe the opposite behavior, where the interference effects are negligible at $T=0$, 
and become important when $T\neq 0$. As recently pointed~\cite{Jalabert:PRL10} out, the effect of a charged 
tip upon $g$ is more important if the QPC is biased outside the conductance plateaus, while the fringes are weaker 
if the QPC is biased inside a plateau. We show in this letter that the temperature can substantially 
{\it{enhance}} the visibility of the fringes, if the QPC is biased near the ends of a plateau. Moreover, the scale 
characterizing the decay of the fringes is not $l_T$, but another length $l_\Gamma$ associated with the sharpness of 
the conductance steps. Fringes persisting above $l_T$ have been seen \cite{Topinka:Nat01}, and the role of impurity 
scattering was believed to be important for explaining this persisting fringing~\cite{Heller:NL05}. Here, we give 
another mechanism for fringes persisting beyond $l_T$, valid without impurity scattering, which takes place at the 
edges of the plateaus and assumes a sharp opening of the QPC conduction channels.   

{\it Numerical observations from QPC models:} 
Our observations are based on numerical simulations of QPC models in the ballistic limit where the only source of 
scattering outside the QPC comes from the depletion region caused by the charged tip. This limit was  
experimentally studied in Refs.~\cite{Jura:natphys,Jura:PRB09}. We have neglected electron-electron interactions 
acting inside the QPC, though they can change the SGM images of a weakly opened contact~\cite{Freyn:PRL08}. 
Therefore, our results will exhibit neither the branches~\cite{Topinka:Nat01}, nor the $0.7  (2e^2/h)$-anomaly seen in 
the experiments. We have used lattice models describing an infinite strip, the QPC being defined in 
a central scattering region. We have taken long adiabatic QPCs for having a sharp opening of 
the conduction channels \cite{Glazman:JETP88}. Model 1 consists in a smooth saddle-point contact~\cite{Buttiker:PRB90}, 
while model 2 has hard walls~\cite{Szafer:PRL89} (see Fig.~\ref{fig1} and its caption). The effect of the charged tip is modelled 
by a site-potential $V\neq 0$ at a distance $r$ from the QPC. We have taken small filling factors for being in the continuum 
limit.

\begin{figure}
\begin{center}
\includegraphics[height=4.4cm]{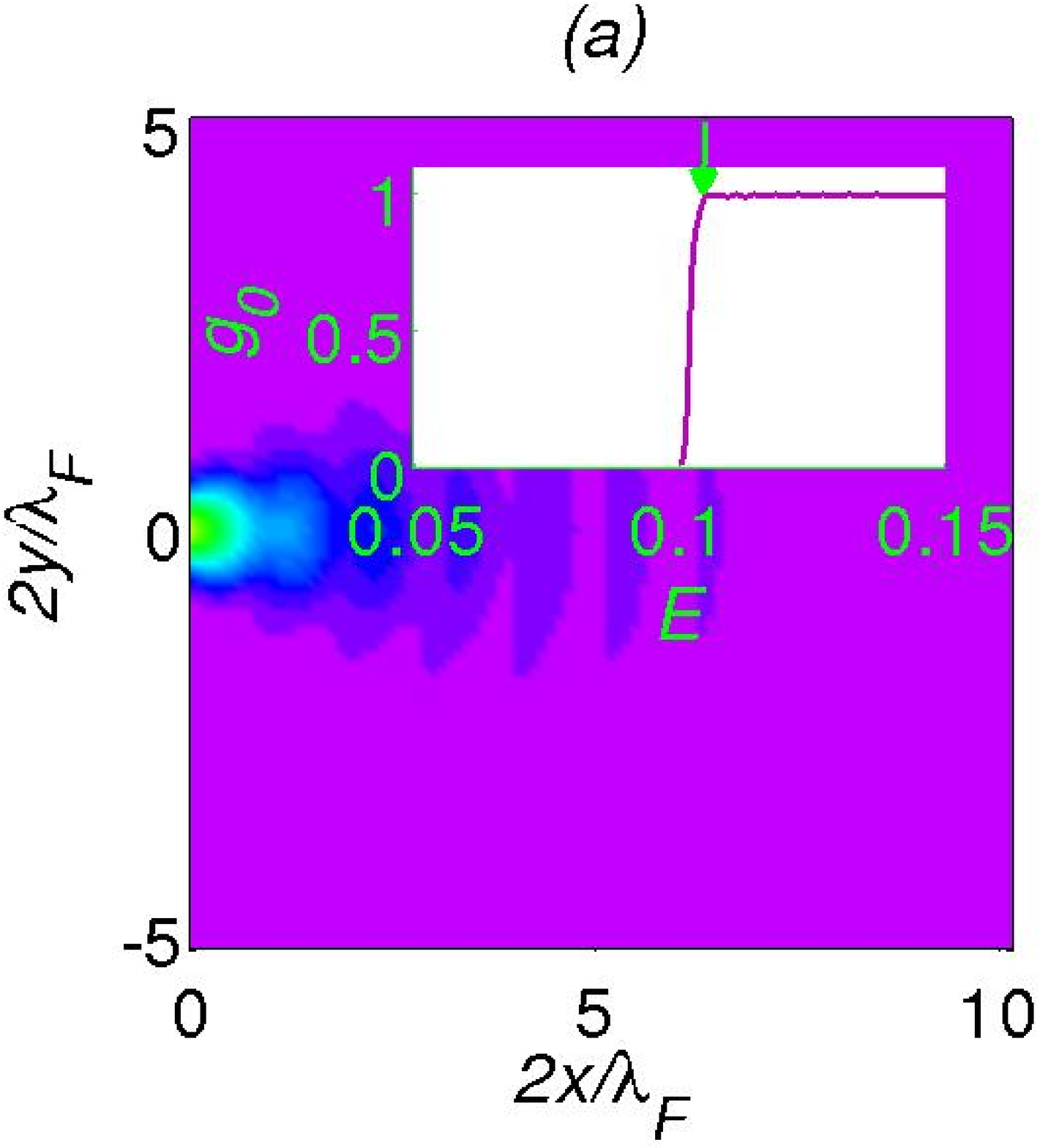}
\includegraphics[height=4.4cm]{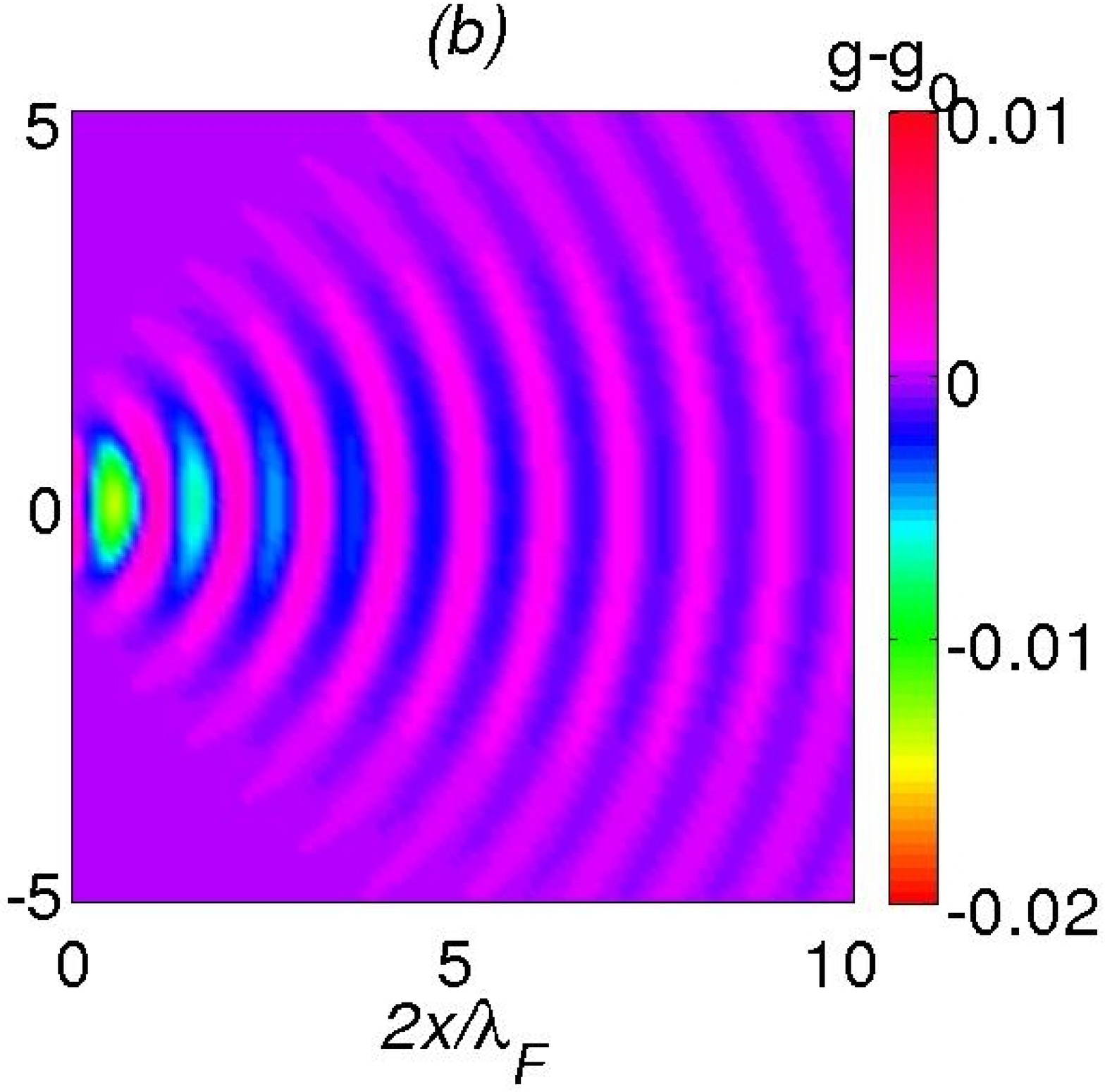}
\end{center}
\begin{center}
\includegraphics[height=7.2cm]{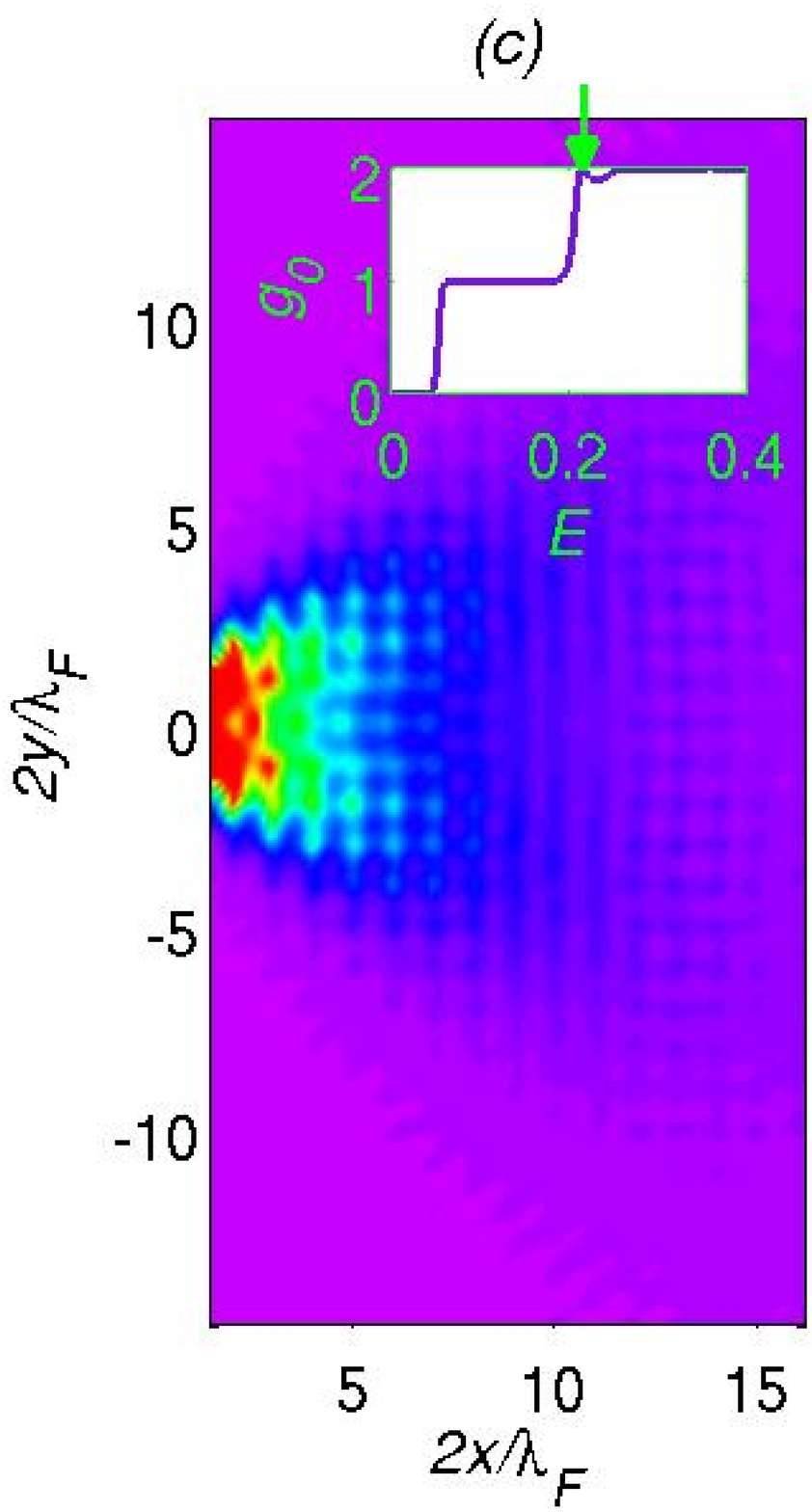}
\includegraphics[height=7.2cm]{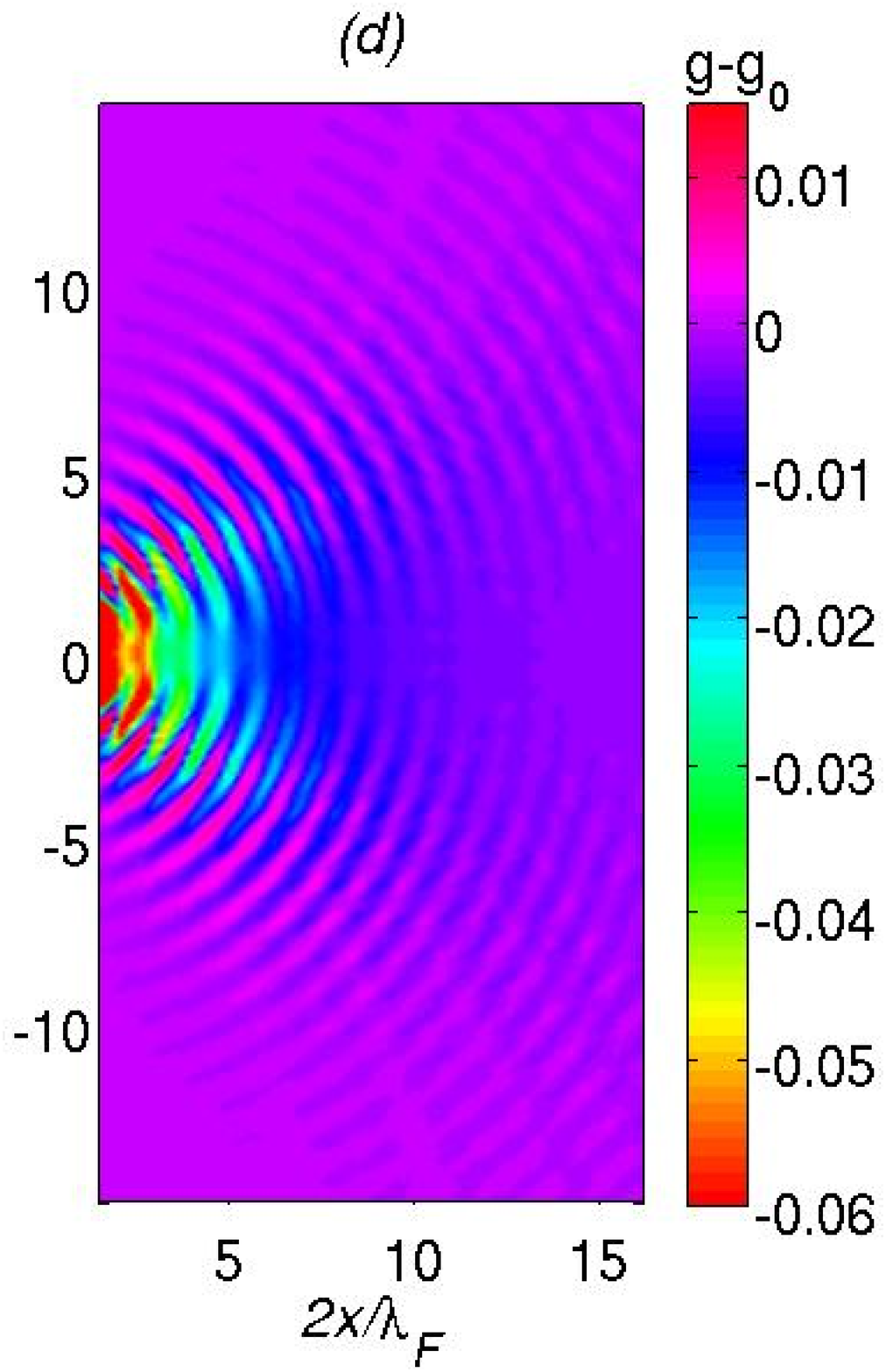}
\end{center}
\caption{\label{fig2}(color online)
$\delta g(T)=g(T)-g_0(T)$ as a function of the tip position (in units 
of $\lambda_F/2$). The left figures correspond to $T=0$, while $T \neq 0$ for 
the right figures. $g_0$ is biased as indicated by the arrow in the insets 
(giving $g_0(T=0)$ as a function of $E_F$). Figs.~\ref{fig1} (a) and (b): 
QPC opened at the beginning of the first plateau using model 1 with $V=1$ 
and $\lambda_F/2=9.65$. $k_BT/E_F=0.01$ for Fig.~\ref{fig1} (b) 
($2l_T/\lambda_F \approx 14.6$). Figs.~\ref{fig1} (c) and (d): 
QPC opened at the beginning of the second  plateau using model 2 with $V=-2$ 
and $\lambda_F/2=6.7$. $k_BT/E_F=0.035$ for Fig.~\ref{fig1} (c) ($2l_T/\lambda_F\approx 4$).
}
\end{figure}

 Typical SGM images are shown in Fig.~\ref{fig2} using a QPC biased at the beginning of the first two conductance plateaus. 
At $T=0$, the conductance without the tip $g_0$ is an integer (see the insets which give $g_0(E)$, the arrows 
indicating the value of $E_F$) and the interference effects are weak [Figs.~\ref{fig2} (a) and (c)]. However, increasing 
$T$ enhances the fringes, as shown in Figs.~\ref{fig2} (b) and (d). When $g_0=1$, the fringes are in the longitudinal $X$ 
direction while they have a $V$ shape when $g_0=2$. This $V$ shape indicates that the thermal enhancement of the fringes 
comes only from a contribution of the second conduction channel, and not of the first. In the middle of the plateaus, 
these angular patterns have been observed in Ref.~\cite{Topinka:Sci00}. A checkerboard pattern of the type discussed in 
Ref.~\cite{Jura:PRB09} can be seen in Fig.~\ref{fig2} (b). Last but not least, in Fig.~\ref{fig2} (d), interference fringes 
can be seen up to $r\approx 4 l_T$ (the thermal length $l_T \approx 4 \lambda_F/2$): thus we have persistent fringing 
beyond $l_T$, a phenomenon which was previously thought to be possible only when there are other scatterers near the tip 
(see \cite{Topinka:Nat01, Heller:NL05}), which is not the case here.
     
{\it Analytical solution of a Resonant Level Model (RLM)}: 
To explain the origin of these temperature induced fringes, we study the simplified interferometer sketched in 
Fig. \ref{fig1} (b) where two semi-infinite square lattices with nearest 
neighbor hopping ($t_h=-1$) are contacted via a single site $\bf{0}$ [of coordinate $(0,0)$] with on-site energy 
$V_{\bf{0}}$ through hopping terms $t_c$. The effect of the tip is modelled by a potential $V\neq 0$ at the site $(x,0)$ 
in the right lead. First, we study the zero temperature limit. 

When $V=0$, the Landauer-B\"uttiker conductance $g_0(E)$ (in units of 
$2(e^2/h)$) is conveniently expressed in terms of the self-energies $\Sigma_{l,r}(E)=R_{l,r}(E)+i I_{l,r}(E)$ 
of the left $l$ and right $r$ leads (Fisher-Lee formula~\cite{Datta:book97}):  
\begin{equation}
g_0(E)= \frac{4 I_{l} I_{r}}
{\left( E - V_{\bf{0}} -R_{l}-R_{r} \right)^2+ 
\left(I_{r}+I_{l}\right)^2 }.
\label{transmission-without-tip}
\end{equation}
$\Sigma_{l,r}(E)$ are related to the retarded Green's functions of the 2 leads evaluated at the 2 sites directly coupled to 
$\bf{0}$: $\Sigma_{l,r}(E)=t_c^2 \left<\pm 1,0|G^R_{l,r}(E)|\pm1,0\right>$. If $t_c$ is small enough, the transmission exhibits  
a Breit-Wigner resonance of width $\Gamma=-(I_l+I_r)$ at an energy $V_{\bf{0}} + R_{l} + R_{r}$. The Green's function 
$G^R_{l,r}(E,V_T=0)$ of semi-infinite square lattices can be obtained from the expression valid for an infinite square 
lattice~\cite{Economou:book06} using the method of mirror images~\cite{Molina:PRB06}. 

In the presence of the tip, the conductance $g(E)$ of the QPC-tip interferometer is still given by Eq.~\eqref{transmission-without-tip}, 
if one adds to the self-energy $\Sigma_{r}(E)$ of the right lead an amount 
$\Delta\Sigma_{r}(E)$ which accounts for the effect of the tip. This generalization of the Fisher-Lee formula uses a  
method introduced in Ref.~\cite{Darancet:PRB10} and will be given in a following paper~\cite{Lemarie:tbp10}. Using Dyson's equation, 
one gets $\Delta\Sigma_{r}(E)=t_c^2 \; \rho\exp(i\phi)\left\vert \left<1,0|G_{r}^R|x,0\right>\right\vert^2$ where $\rho$ and $\phi$ 
are respectively the modulus and the argument of the scattering amplitude $V/(1-V\left<x,0|G_{r}^R|x,0\right>)$. Hereafter, the 
$x$-dependences of $\rho$ and $\phi$ are neglected, an assumption valid if $x$ is sufficiently large. In the continuum limit (i.e momentum  
$k \ll 1$ and energy $E\approx k^2$) and at large distance $k x \gg 1$:
\begin{eqnarray}
\frac{\Delta \Sigma_r(E)}{t_c^2} \approx -\dfrac{{\rho k \exp[i(2 k x + \pi/2 + {\phi})]}}{2 \pi x} 
+ O\left(\dfrac{1}{{x}^{3/2}}\right) .
\label{correction-self-energy}
\end{eqnarray}
Since $\Delta\Sigma_{r}(E) \to 0$ as $x$ increases, the effect $\delta g=g-g_0$ of the tip upon 
$g_0$ can be expanded in powers of the reduced variables $\delta R=\Delta R/I$ and 
$\delta I= \Delta I/I$ (with $I=I_{r,l}\approx -t_c^2 k^2/4$ and $\Delta\Sigma_{r}=\Delta R+i \Delta I$).
The coefficients depend on $g_0$ and on $S_O=g_0(1-g_0)$, the QPC shot noise~\cite{PhysTod:2003}:
\begin{eqnarray}
\begin{aligned}
\delta g=& s g_0 \sqrt{S_0} \; \delta R + S_0 \; \delta I + s g_0\sqrt{S_0} (1-2 g_0)\; \delta R \delta I\\
&+ 
g_0^2 \left( \dfrac{3}{4} - g_0\right)\; \delta R^2 + g_0^2 \left(-\dfrac{5}{4}  + 
g_0\right) \;\delta I^2 \; ,
\label{expansion}
\end{aligned}
\end{eqnarray}
where $s=\text{sign}\left[(E-V_{\bf{0}}-2R)/2I\right]$. 

Out of resonance ($g_0 < 1$), the linear terms give a large oscillatory effect of the tip with period $\lambda_F/2$ 
and $1/x$-decay:
\begin{equation}\label{eq:deltagg0neq1}
\dfrac{\delta {g}}{g_0} \underset{g_0 <1}{\approx}  \dfrac{2 {\rho\sin \zeta_0}\cos (2 k_F x + 
\theta)}{\pi k_F x }\; + O\left( \dfrac{1}{{x}^{3/2}}\right)\; ,
\end{equation}
where $\theta \equiv \pi/2 + {\phi} - \zeta_0$ with $\sin \zeta_0 \equiv s \sqrt{1-{g}_0}$. 
At resonance ($g_0= 1$), the linear terms vanish and the quadratic terms give a non oscillatory negative 
correction which falls off as $1/x^2$ accompanied by an oscillatory term $\delta g_{\text{osc}}$ with period
$\lambda_F/2$ and $1/x^{5/2}$-decay:
 \begin{equation}\label{eq:deltagg01}
\dfrac{\delta {g}}{g_0} \underset{g_0=1}{\approx} - \left(\dfrac{{\rho}}{\pi k_F x}\right)^2 - \delta {g}_{\text{osc}}\; .
\end{equation}
This suppression of the linear terms at resonance for the RLM model and the perturbative result derived in 
Ref.~\cite{Jalabert:PRL10} for a QPC yield the same conclusion: interfering electrons are mainly those which 
contribute to the shot noise $S_0$, those of energy around the resonance for the RLM model and those of energy 
between the plateaus for the QPC. 

\begin{figure}
\centerline{
\includegraphics[width=\linewidth]{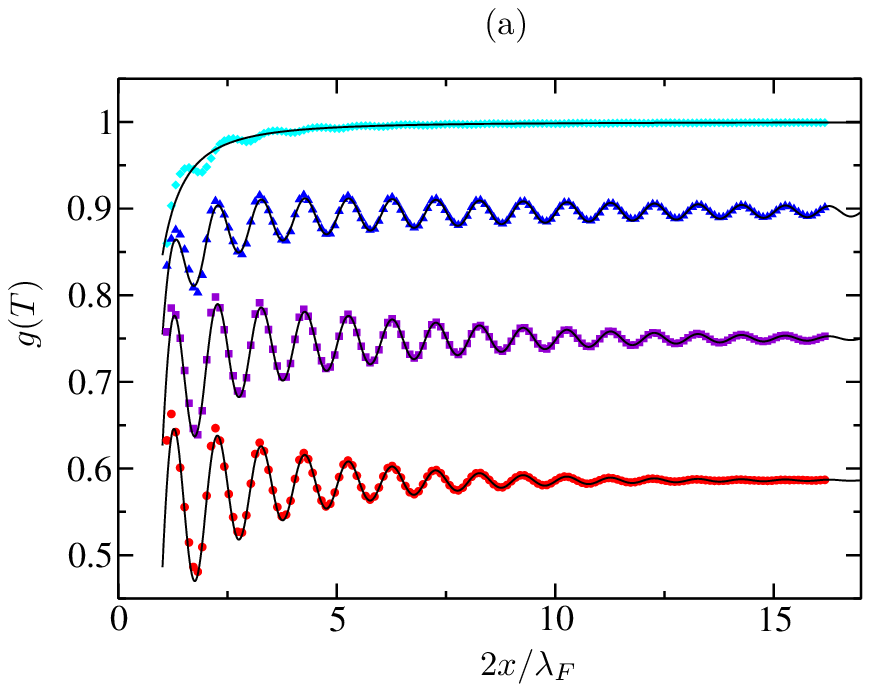}}
\centerline{
\includegraphics[width=\linewidth]{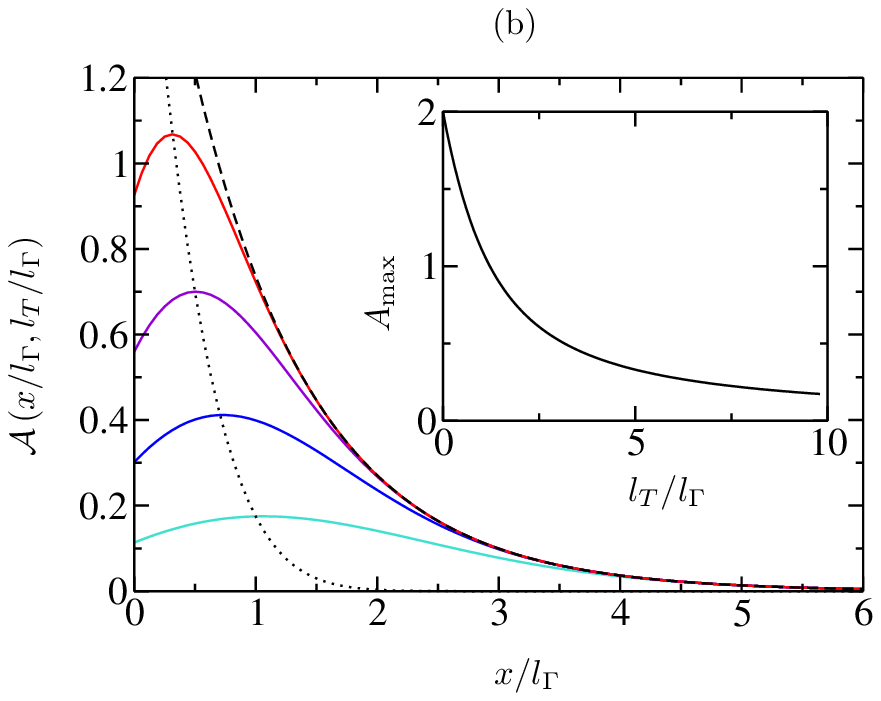} 
}
\caption{\label{fig3} (Color online) (a) RLM conductance $g(T)$ as a function of $2x/\lambda_F$ for 
increasing temperatures (from top to bottom). The color points are results obtained by direct numerical 
simulations, taking from top to bottom $2l_T/\lambda_F= \infty$, $8.8$, $4.4$ and $2.2$. The solid lines 
give the analytical results (Eq.~\eqref{delta-temperature}). 
$t_c=0.4$, $V=-2$, $\lambda_F/2 \approx 10$ and $2 l_\Gamma/\lambda_F \approx 4$. 
(b) As a function of $x/l_\Gamma$, rescaled amplitude ${\cal A}=A\left(x/l_\Gamma,l_T/l_\Gamma\right) 
\text{erfc}(l_T/l_\Gamma)$ for fixed $l_{\Gamma}$ and increasing temperatures, i.e.
ratios $l_T/l_\Gamma=1$ (light blue), $0.7$ (blue), $0.5$ (violet) and $0.3$ (red) (solid lines from bottom to top). 
The dashed and dotted lines give respectively the asymptotic behavior $2 \exp (-x/l_\Gamma)$ valid for 
$x \gg l_T$ and the function ${\cal A}\left(l_T/l_\Gamma,l_T/l_\Gamma\right)$ where 
${\cal A}$ is maximum. Inset: maximum $A_{\text{max}}$ of the bare amplitude $A$ as a function of $l_T/l_\Gamma$. 
} 
\end{figure}

Let us now consider the temperature dependence of these interferences when $E_F$ is located on the transmission 
peak ($g_0=1$ when $T=0$). The quantum statistics give an energy scale $\approx k_BT$. The resonant contact 
gives another energy scale, since it restricts the transmission inside an energy window $\Gamma$ around $E_F$. 
This gives two length scales (over which an electron propagates at the Fermi velocity 
during the associated time scales) the thermal length $l_T=k_F/(4 k_B T \pi^{-1/2})$ yielded by the quantum statistics and 
the length $l_\Gamma=k_F/\Gamma$ yielded by the resonance. The temperature dependence of the fringes 
is a function of these two scales. The conductance (in units of $2(e^2/h)$) of the contact 
at a temperature $T$ reads:
\begin{equation}\label{temperature}
g_0(T)=\int_{0}^{\infty} g_0(E) \left[-\frac{\partial f_{T}(E)}{\partial E}\right] dE\; ,
\end{equation}
the conductance $g_0(E)$ at $T=0$ characterizing the transmission of an electron of energy $E$ through 
the contact. The derivative of $f_{T}(E)$ is approximately given by $-4k_B T \partial f_{T}(E)/\partial E 
\approx \exp-[(E-E_F)/(4 k_B T \pi^{-1/2})]^2$. The effect of the tip $\delta g(T)$ upon $g_0(T)$ is given 
by Eq.~\eqref{temperature}, taking the change $\delta g(E)$ (Eq.~\eqref{expansion}) at $T=0$ 
instead of $g_0(E)$. At resonance and $T=0$, $\delta g(x)$ shows only very weak 
oscillations as $x$ varies (Eq. \eqref{eq:deltagg01}). If $T\neq 0$, the electrons of non resonant energies 
$E$ around $E_F$ enhance the interference effects via their linear $\rho/x$ contributions in the expansion 
\eqref{expansion}, which become non zero when $S_0(E)\neq0$. However, the thermal enhancement 
of the fringes at short distances vanishes at long distances, since the fast oscillations of the linear terms 
as $E$ varies destroy the interferences when $k x \gg 1$. To check this quantitatively, 
one must calculate the integral over energy explicitly. Doing standard approximations (see Ref.~\cite{Gruner:PRL72}), 
we get:
\begin{equation}\label{delta-temperature}
\frac{\delta g(T)}{g_0(T)} \approx  A\left(\frac{x}{l_\Gamma},\frac{l_T}{l_\Gamma}\right) \frac{\rho 
\cos\left(2k_Fx+\phi \right)}{2\pi k_F x} -\dfrac{{\rho}^2}{(\pi k_F x)^2}.
\end{equation}
The amplitude $A$ (shown in Fig.~\ref{fig3} (b)) is given by 
\begin{equation}\label{amplitude}
A\left(\mu,\nu\right)=\frac{(1+2 \mu+4 \nu^2)F_{+}+F_{-}-G}{\text{erfc}(\nu)},
\end{equation}
where $F_{\pm}=e^{\pm \mu} \text{erfc} \left[\frac{2\nu^2 \pm \mu}{2\nu}\right]$ and $G=\frac{4 \nu}{\sqrt{\pi}} 
e^{- \nu^2-\frac{\mu^2}{4\nu^2}}$. Three main results can be seen: (i) $A$ vanishes when $T \rightarrow 0$, i.e. 
when $l_T \gg l_\Gamma$; (ii) When $x \gg l_T$, there is a universal asymptotic $\exp (-x/l_{\Gamma})$-decay characterized by 
the $T$-independent scale $l_\Gamma$, and not by $l_T$; 
(iii) $A$ has a maximum when $x \approx l_T$. Fig.~\ref{fig3} (a) shows a comparison between the analytical 
formula \eqref{delta-temperature} and the results of numerical simulations of the RLM. The agreement is very good, 
without any adjustable parameter. 

For the RLM model, $g_0(E)$ is given by a Lorentzian and $g_0=1$ only at resonance. For a QPC, $g_0(E)$ is a step-like 
function where $g_0=1$ over the first plateau. For extending the results of the RLM model to a QPC with $g_0 \leq 1$, 
we summarize the arguments which will be given in a following paper~\cite{Lemarie:tbp10}. Firstly, we have checked 
that the expansion ~\eqref{expansion} describes also (up to a phase factor) the effect of a tip upon a QPC when 
$g_0 \leq 1$, if one uses in Eq.~\eqref{expansion} the step-like function $g_0(E)$ of the QPC instead of the Lorentzian 
$g_0(E)$ of the RLM model. Secondly, the linear terms in Eq.~\eqref{expansion} depend on $g_0(E)$ through the combination 
$g_0(E)[1-g_0(E)]$ which characterizes the QPC shot noise. For a QPC, the scale $\Gamma$ is thus given by the energy scale 
over which the QPC shot noise (and not the QPC transmission) is important. If $l_\Gamma \gg l_T$, persistent fringing beyond 
$l_T$ can be observed even without impurity scattering.
 
 In summary, the SGM fringes observed at low temperatures using a QPC biased near the ends of a conductance plateau 
are mainly due to a thermal effect which vanishes when $T \to 0$ and decays with a $T$-independent length $l_\Gamma$. 
More generally, the thermal enhancement and the persistence beyond $l_T$ of interference effects can be observed  
if one uses electron interferometers made of two scatterers, one of them having a resonance at $E_F$. Fermi-Dirac 
statistics give rise to a temperature induced interferometer which disappears as $T \to 0$ if one of the 
scatterers is transparent at $E_F$ and is only seen by the electrons of energy $E \neq E_F$. Moreover, the resonant 
scatterer acts as a filter, yielding interferences over a scale independent of the temperature (a somewhat related 
phenomenon has been observed in Ref~\cite{Molenkamp:PRL95}). Studying the SGM images, we have shown that a QPC 
also can filter the interfering electrons as does a resonant scatterer, allowing better quantum interference effects.

 We thank A. Freyn, R. A. Jalabert, K. A. Muttalib, F. Portier and D. Weinmann for useful discussions, and the French National 
Agency ANR (project ANR-08-BLAN-0030-02 ``Item-Th'') for financial support. 


\begin{thebibliography}{24}
\expandafter\ifx\csname natexlab\endcsname\relax\def\natexlab#1{#1}\fi
\expandafter\ifx\csname bibnamefont\endcsname\relax
  \def\bibnamefont#1{#1}\fi
\expandafter\ifx\csname bibfnamefont\endcsname\relax
  \def\bibfnamefont#1{#1}\fi
\expandafter\ifx\csname citenamefont\endcsname\relax
  \def\citenamefont#1{#1}\fi
\expandafter\ifx\csname url\endcsname\relax
  \def\url#1{\texttt{#1}}\fi
\expandafter\ifx\csname urlprefix\endcsname\relax\def\urlprefix{URL }\fi
\providecommand{\bibinfo}[2]{#2}
\providecommand{\eprint}[2][]{\url{#2}}

\bibitem{PhysRevLett.60.848} B.~J.~van Wees {\it et~al.}, Phys. Rev. Lett. {\bf 60}, 848 (1988).

\bibitem{Wharam} D.~Wharam {\it et~al.}, J. Phys. C {\bf 21}, L209 (1988).

\bibitem[{\citenamefont{Glazman et~al.}(1988)\citenamefont{Glazman, Lesovik,
  Kmelnitskii, and Shekhter}}]{Glazman:JETP88}
\bibinfo{author}{\bibfnamefont{L.}~\bibnamefont{Glazman}},
  \bibinfo{author}{\bibfnamefont{G.}~\bibnamefont{Lesovik}},
  \bibinfo{author}{\bibfnamefont{D.}~\bibnamefont{Kmelnitskii}},
  \bibnamefont{and} \bibinfo{author}{\bibfnamefont{R.}~\bibnamefont{Shekhter}},
  \bibinfo{journal}{JETP Lett.} \textbf{\bibinfo{volume}{48}},
  \bibinfo{pages}{238} (\bibinfo{year}{1988}).

\bibitem[{\citenamefont{B\"uttiker}(1990)}]{Buttiker:PRB90}
\bibinfo{author}{\bibfnamefont{M.}~\bibnamefont{B\"uttiker}},
  \bibinfo{journal}{Phys. Rev. B} \textbf{\bibinfo{volume}{41}},
  \bibinfo{pages}{7906} (\bibinfo{year}{1990}).

\bibitem[{\citenamefont{Beenakker and Van~Houten}(1991)}]{Beenakker:SSP91}
\bibinfo{author}{\bibfnamefont{C.~W.~J.} \bibnamefont{Beenakker}}
  \bibnamefont{and}
  \bibinfo{author}{\bibfnamefont{H.}~\bibnamefont{Van~Houten}},
  \bibinfo{journal}{Solid State Physics} \textbf{\bibinfo{volume}{44}},
  \bibinfo{pages}{1} (\bibinfo{year}{1991}),
  \urlprefix\url{arXiv:cond-mat/0412664v1}.

\bibitem{PhysRevLett.77.135} K.~J.~Thomas {\it et~al.}, Phys. Rev. Lett. {\bf 77}, 135 (1996).

\bibitem[{\citenamefont{Topinka et~al.}(2000)\citenamefont{Topinka, LeRoy,
  Shaw, Heller, Westervelt, Maranowski, and Gossard}}]{Topinka:Sci00}
\bibinfo{author}{\bibfnamefont{M.}~\bibnamefont{Topinka}},
  \bibinfo{author}{\bibfnamefont{B.}~\bibnamefont{LeRoy}},
  \bibinfo{author}{\bibfnamefont{S.}~\bibnamefont{Shaw}},
  \bibinfo{author}{\bibfnamefont{E.}~\bibnamefont{Heller}},
  \bibinfo{author}{\bibfnamefont{R.}~\bibnamefont{Westervelt}},
  \bibinfo{author}{\bibfnamefont{K.}~\bibnamefont{Maranowski}},
  \bibnamefont{and} \bibinfo{author}{\bibfnamefont{A.}~\bibnamefont{Gossard}},
  \bibinfo{journal}{Science} \textbf{\bibinfo{volume}{289}},
  \bibinfo{pages}{2323} (\bibinfo{year}{2000}).

\bibitem[{\citenamefont{Topinka et~al.}(2001)\citenamefont{Topinka, LeRoy,
  Westervelt, Shaw, Fleischmann, Heller, Maranowski, and
  Gossard}}]{Topinka:Nat01}
\bibinfo{author}{\bibfnamefont{M.}~\bibnamefont{Topinka}},
  \bibinfo{author}{\bibfnamefont{B.}~\bibnamefont{LeRoy}},
  \bibinfo{author}{\bibfnamefont{R.}~\bibnamefont{Westervelt}},
  \bibinfo{author}{\bibfnamefont{S.}~\bibnamefont{Shaw}},
  \bibinfo{author}{\bibfnamefont{R.}~\bibnamefont{Fleischmann}},
  \bibinfo{author}{\bibfnamefont{E.}~\bibnamefont{Heller}},
  \bibinfo{author}{\bibfnamefont{K.}~\bibnamefont{Maranowski}},
  \bibnamefont{and} \bibinfo{author}{\bibfnamefont{A.}~\bibnamefont{Gossard}},
  \bibinfo{journal}{Nature} \textbf{\bibinfo{volume}{410}},
  \bibinfo{pages}{183} (\bibinfo{year}{2001}).

\bibitem[{\citenamefont{Topinka et~al.}(2003)\citenamefont{Topinka, Westervelt,
  and Heller}}]{Topinka:PT03}
\bibinfo{author}{\bibfnamefont{M.}~\bibnamefont{Topinka}},
  \bibinfo{author}{\bibfnamefont{R.}~\bibnamefont{Westervelt}},
  \bibnamefont{and} \bibinfo{author}{\bibfnamefont{E.}~\bibnamefont{Heller}},
  \bibinfo{journal}{Physics Today} \textbf{\bibinfo{volume}{56}},
  \bibinfo{pages}{47} (\bibinfo{year}{2003}).
\bibitem{LeRoy:PRL05} B.~J.~LeRoy {\it et~al.}, Phys. Rev. Lett. {\bf 94}, 126801 (2005).

\bibitem[{\citenamefont{Jalabert et~al.}(2010)\citenamefont{Jalabert, Szewc,
  Tomsovic, and Weinmann}}]{Jalabert:PRL10}
\bibinfo{author}{\bibfnamefont{R.~A.} \bibnamefont{Jalabert}},
  \bibinfo{author}{\bibfnamefont{W.}~\bibnamefont{Szewc}},
  \bibinfo{author}{\bibfnamefont{S.}~\bibnamefont{Tomsovic}}, \bibnamefont{and}
  \bibinfo{author}{\bibfnamefont{D.}~\bibnamefont{Weinmann}},
  \bibinfo{journal}{Phys. Rev. Lett.} \textbf{\bibinfo{volume}{105}},
  \bibinfo{pages}{166802} (\bibinfo{year}{2010}).


\bibitem{Jura:natphys} M.~P.~Jura {\it et~al.}, Nat. Phys. {\bf 3}, 841 (2007).

\bibitem[{\citenamefont{Jura et~al.}(2009)\citenamefont{Jura, Topinka, Grobis,
  Pfeiffer, West, and Goldhaber-Gordon}}]{Jura:PRB09}
\bibinfo{author}{\bibfnamefont{M.~P.} \bibnamefont{Jura}},
  \bibinfo{author}{\bibfnamefont{M.~A.} \bibnamefont{Topinka}},
  \bibinfo{author}{\bibfnamefont{M.}~\bibnamefont{Grobis}},
  \bibinfo{author}{\bibfnamefont{L.~N.} \bibnamefont{Pfeiffer}},
  \bibinfo{author}{\bibfnamefont{K.~W.} \bibnamefont{West}}, \bibnamefont{and}
  \bibinfo{author}{\bibfnamefont{D.}~\bibnamefont{Goldhaber-Gordon}},
  \bibinfo{journal}{Phys. Rev. B} \textbf{\bibinfo{volume}{80}},
  \bibinfo{pages}{041303} (\bibinfo{year}{2009}).


\bibitem[{\citenamefont{Heller et~al.}(2005)\citenamefont{Heller, Aidala,
  LeRoy, Bleszynski, Kalben, Westervelt, Maranowski, and
  Gossard}}]{Heller:NL05}
\bibinfo{author}{\bibfnamefont{E.}~\bibnamefont{Heller}},
  \bibinfo{author}{\bibfnamefont{K.}~\bibnamefont{Aidala}},
  \bibinfo{author}{\bibfnamefont{B.}~\bibnamefont{LeRoy}},
  \bibinfo{author}{\bibfnamefont{A.}~\bibnamefont{Bleszynski}},
  \bibinfo{author}{\bibfnamefont{A.}~\bibnamefont{Kalben}},
  \bibinfo{author}{\bibfnamefont{R.}~\bibnamefont{Westervelt}},
  \bibinfo{author}{\bibfnamefont{K.}~\bibnamefont{Maranowski}},
  \bibnamefont{and} \bibinfo{author}{\bibfnamefont{A.}~\bibnamefont{Gossard}},
  \bibinfo{journal}{Nano Lett.} \textbf{\bibinfo{volume}{5}},
  \bibinfo{pages}{1285} (\bibinfo{year}{2005}).

\bibitem[{\citenamefont{Freyn et~al.}(2008)\citenamefont{Freyn, Kleftogiannis,
  and Pichard}}]{Freyn:PRL08}
\bibinfo{author}{\bibfnamefont{A.}~\bibnamefont{Freyn}},
  \bibinfo{author}{\bibfnamefont{I.}~\bibnamefont{Kleftogiannis}},
  \bibnamefont{and} \bibinfo{author}{\bibfnamefont{J.-L.}
  \bibnamefont{Pichard}}, \bibinfo{journal}{Phys. Rev. Lett.}
  \textbf{\bibinfo{volume}{100}}, \bibinfo{pages}{226802}
  (\bibinfo{year}{2008}).

\bibitem[{\citenamefont{Szafer and Stone}(1989)}]{Szafer:PRL89}
\bibinfo{author}{\bibfnamefont{A.}~\bibnamefont{Szafer}} \bibnamefont{and}
  \bibinfo{author}{\bibfnamefont{A.~D.} \bibnamefont{Stone}},
  \bibinfo{journal}{Phys. Rev. Lett.} \textbf{\bibinfo{volume}{62}},
  \bibinfo{pages}{300} (\bibinfo{year}{1989}).

\bibitem[{\citenamefont{Datta}(1997)}]{Datta:book97}
\bibinfo{author}{\bibfnamefont{S.}~\bibnamefont{Datta}},
  \emph{\bibinfo{title}{{Electronic transport in mesoscopic systems}}}
  (\bibinfo{publisher}{Cambridge Univ Pr}, \bibinfo{year}{1997}).

\bibitem[{\citenamefont{Economou}(2006)}]{Economou:book06}
\bibinfo{author}{\bibfnamefont{E.}~\bibnamefont{Economou}},
  \emph{\bibinfo{title}{{Green's functions in quantum physics}}}
  (\bibinfo{publisher}{Springer Verlag}, \bibinfo{year}{2006}).

\bibitem[{\citenamefont{Molina}(2006)}]{Molina:PRB06}
\bibinfo{author}{\bibfnamefont{M.~I.} \bibnamefont{Molina}},
  \bibinfo{journal}{Phys. Rev. B} \textbf{\bibinfo{volume}{74}},
  \bibinfo{pages}{045412} (\bibinfo{year}{2006}).

\bibitem[{\citenamefont{Darancet et~al.}(2010)\citenamefont{Darancet, Olevano,
  and Mayou}}]{Darancet:PRB10}
\bibinfo{author}{\bibfnamefont{P.}~\bibnamefont{Darancet}},
  \bibinfo{author}{\bibfnamefont{V.}~\bibnamefont{Olevano}}, \bibnamefont{and}
  \bibinfo{author}{\bibfnamefont{D.}~\bibnamefont{Mayou}},
  \bibinfo{journal}{Phys. Rev. B} \textbf{\bibinfo{volume}{81}},
  \bibinfo{pages}{155422} (\bibinfo{year}{2010}).

\bibitem[{\citenamefont{Lemari\'e et~al.}(2010)\citenamefont{Lemari\'e, Abbout,
  and Pichard}}]{Lemarie:tbp10}
\bibinfo{author}{\bibfnamefont{G.}~\bibnamefont{Lemari\'e}},
  \bibinfo{author}{\bibfnamefont{A.}~\bibnamefont{Abbout}}, \bibnamefont{and}
  \bibinfo{author}{\bibfnamefont{J.-L.} \bibnamefont{Pichard}},
  \bibinfo{journal}{in preparation}.

\bibitem[{\citenamefont{Beenakker and Sch\"onenberger}(May
  2003)}]{PhysTod:2003}
\bibinfo{author}{\bibfnamefont{C.~W.~J.} \bibnamefont{Beenakker}}
  \bibnamefont{and}
  \bibinfo{author}{\bibfnamefont{C.}~\bibnamefont{Sch\"onenberger}},
  \bibinfo{journal}{Physics Today} p.~\bibinfo{pages}{37} (\bibinfo{year}{May
  2003}).

\bibitem[{\citenamefont{Mezei and Gr\"uner}(1972)}]{Gruner:PRL72}
\bibinfo{author}{\bibfnamefont{F.}~\bibnamefont{Mezei}} \bibnamefont{and}
  \bibinfo{author}{\bibfnamefont{G.}~\bibnamefont{Gr\"uner}},
  \bibinfo{journal}{Phys. Rev. Lett.} \textbf{\bibinfo{volume}{29}},
  \bibinfo{pages}{1465} (\bibinfo{year}{1972}).

\bibitem{Molenkamp:PRL95} N.~C.~van~der Vaart {\it et~al.}, Phys. Rev. Lett. {\bf 74}, 4702 (1995).

\end{thebibliography}

\end{document}